\documentclass[10pt,a4paper,twocolumn,showkeys,showpacs]{revtex4-2}
\usepackage{mathptm} 
\usepackage{graphicx}
\usepackage{color}
\begin{document}

\title{Information theoretic measures on quantum droplets in ultracold atomic systems }

\author{Sk Siddik and Golam Ali Sekh }\email{skgolamali@gmail.com}
\affiliation{Department of Physics, Kazi Nazrul University, Asansol-713340, W.B., India  }

\begin{abstract}
We consider Shannon entropy, Fisher information, R\'enyi entropy, and Tsallis entropy to study the quantum droplet  phase in  Bose-Einstein condensates. In the beyond mean-field description, the Gross-Pitaevskii equation with Lee-Huang-Yang correction gives a family of quantum droplets with different chemical potentials.  At a larger value of chemical potential, quantum droplet with sharp-top probability density distribution starts to form  while it becomes flat top for a smaller value of chemical potential.   We  show that  entropic measures can  distinguish  the shape change of the probability density distributions and thus can identify the onset of the droplet phase. During the onset of droplet phase, the Shannon entropy decreases gradually  with the decrease of chemical potential and attains a minimum in the vicinity where a smooth transition from flat-top to sharp-top QDs occurs. At this stage, the Shannon entropy increases abruptly with the lowering of chemical potential. We observe an opposite trend in the case of Fisher information. These results are found to be consistent with the R\'enyi and Tsallis entropic measures.     
		  
\end{abstract} 
\maketitle 
\section{Introduction}   
Besides the mean-field interaction, one can expect another type of interaction due to quantum fluctuations in the Bose Einstein condensates(BECs). The interplay between quantum fluctuation  due to Lee-Huang-Yang (LHY) correction and  effective mean-field interaction allows the formation of self-bound liquid like states or the so- called quantum droplets(QDs) \cite{r0}. Generation of QDs is purely a manifestation of quantum nature of BECs \cite{r3}. QDs exhibit intriguing properties of quantum mechanics at a macroscopic scale, and allows to study fundamental quantum phenomena. It may  potentially pave the way for applications in the field like precision measurement, quantum computing and also in understanding super-fluidity  and superconductivity \cite{q7,q8}.  
	  
After the theoretical prediction of QDs in 2015 \cite{r0} and its first  experimental realization in 2018 \cite{r01}, studies on QDs in ultra-cold atomic system have received a tremendous momentum both theoretically and experimentally. The QDs have been observed experimentally both in dipolar BECs and non- dipolar  Bose-Bose mixtures \cite{f2}. It is seen that the self-bound immicible droplets in dipolar mixture lead to the formation of droplet molecules due to inter-component interaction \cite{f1}. Current theoretical works on this topic include the existence of QDs in Bose-Fermi mixtures \cite{r1,r2},  generation of supersolid crystals with QDs \cite{f3} and investigation on the properties of QDs in periodic potentials \cite{p4,p5,p6}. It is worth mentioning that the interplay between attractive mean-field interaction and group velocity dispersion produces a self-reinforce localized  matter-wave, called soliton, in BECs if the quantum fluctuation is negligible \cite{n1}. However,  a transition may occur with the inclusion of LHY correction from soliton to quantum droplet  if the  mean-field attractive interaction and transverse confinement are appropriately chosen {\cite{s2,s3,sr1,sr2}}.   
	
Our objective in this  paper is to  study   quantum droplets phase with the help of  (i) Shannon entropy, (ii) Fisher information, (iii) R\'enyi entropy, (iv) Tsallis entropy. These information theoretic measures differ in their ability to identify characteristics of the system.  For example, the Shannon entropy is sensitive to the global changes of probability density distributions whereas the Fisher information is sensitive to the local changes of the distributions in a small-sized region \cite{r4,r5}. In the statistical point of view, Fisher information can be realized as a measure of disorder or smoothness of a probability density.  These two information measures are  complementary to each other and efficient to extract information from many complicated systems. In the recent past, these two information measures are used to study  Morse oscillator  \cite{ss1}, stripe phase in spin-orbit coupled BECs \cite{g1,g2}, localization and delocalization of matter waves \cite{gs1,gs2}, properties of K -shell electron of neutral atoms \cite{ss01,ssg1,n2} and, revival and collapse of quantum wave packets\cite{ss2}. R\'enyi entropy can efficiently deal with quantum entanglement\cite{r6}, thermodynamic stability of black hole \cite{r7}, quantum phase transition in Dicke model \cite{r8}. It is also found as extensive for the Laplace-de Finetti distribution\cite{r9}. {Recently,} a  similar method has been used to study the transition from sharp to flat-top optical solitons in cubic-quintic nonlinear media \cite{m2}. Tsallis entropy is  very effective to investigate non extensive effects in self-gravitating systems and in heterogeneous plasma \cite{r05,r03,r04,m1}. In the limiting case, however, both R\'enyi and Tsallis entropies approach to the Shannon entropy.

In quasi-one-dimensional geometry, QDs supported by the Gross-Pitaevskii equation  with LHY correction  have sharp-top density distributions at the onset of QD phase. It is associated with a larger value of chemical potential. However, sharpness of the top decreases with the decrease of chemical potential and the top of the distribution becomes sufficiently flat at a larger value of chemical potential.  In this work, we see that the information  theoretic measures are very effective  to detect the change in the characteristics of quantum droplets(QDs). Particularly, Shannon entropy rapidly increases while Fisher information decreases for the flat-top QDs. This nature of variation remains same  near the  threshold of the droplet phase where QDs have sharp-top density distributions.  However, their increment is abrupt for flat-top QDs while it is regular for sharp-top QDs.  In between of these two limiting cases, we see that the Shannon entropy  attains its minimum value while Fisher information becomes maximum in coordinate space. Other two information measures, R\'enyi and Tsallis entropies,  also vary with shape change of QDs but the change in Tsallis entropy is smaller than that in R\'enyi entropy.   
	
In the section II, we introduce  the theoretical model to describe BECs in presence of  quantum fluctuation. More specifically, we consider a Gross-Pitaevskii equation (GPE) \cite{q6} with Lee-Huang-Yang term and study the probability density distribution of the system in the vicinity of quantum droplet phase both in coordinate and momentum spaces. In section III, we present the results of information theoretic measures for different values of chemical potentials and also check the associated entropic uncertainty relations. {We also introduce Shannon-Fisher complexity and show that its value is larger for a flat-top QD than that for a sharp-top QD}. In section IV, we make concluding remarks.

\section{Theoretical formulation for quantum droplets} 
Bose-Einstein condensate (BEC) is a state of matter in which  a macroscopically large number of particles tends to coalesce into the lowest  energy state as the temperature approaches to the critical value for the formation of BECs. It is a weakly interacting system where  the quantum fluctuations are negligible at a temperature $\left( T\right) $ much below the critical temperature $\left( T_{c}\right) $. Theoretically, the condensate $\left( T<T_{c}\right) $ is described by the  mean-field Gross-Piteavskii equation (GPE) containing the effects of inter-atomic interaction.  However, in Bose- Bose mixtures some situations arise, where quantum fluctuations are not negligible, and  thus the correction due to  Lee-Huang-Yang (LHY) needs to be incorporated\cite{r02}. Particularly, the fine tuning between attractive and repulsive atomic  interaction  leads to the formation of quantum droplets (QDs) for $\delta g>0$. Here, $\delta g=g_{12}+\sqrt{g_{11}g_{22}}$ with $g_{ii}$ and $g_{12}$ are intra-atomic and  inter-atomic interaction respectively.  Considering equal atomic populations of the binary mixture components ($N_{1}=N_{2}$) and $g_{11}=g_{22}=g$, one can derive an effective one dimensional(1-D) equation, often called, effective Gross-Piteavskii equation (eGPE). The eGPE  is given by  \cite{r10}. 
\begin{equation} 
i\hbar\frac{\partial\psi}{\partial t}=-\frac{\hbar^{2}}{2m}\frac{\partial^{2}\psi}{\partial x^{2}}+V\left(x \right)\psi  	+\delta g|\psi|^{2}\psi-\Omega|\psi|\psi.
\label{Eq1}
\end{equation}  
Here $V\left(x \right) $ is the external trapping potential and  $\Omega=\frac{\sqrt{2m}}{\pi\hbar}g^{\frac{3}{2}}$. Understandably, the eGPE describes beyond mean-field effect through the last term in Eq.(\ref{Eq1}). For convenience,  we define the following dimensionless variables  
\begin{equation}
\tilde{t}=\frac{t}{t_{0}},\,\,\, \tilde{x}=\frac{x}{x_{0}},\,\,\, \tilde{\psi}=\frac{\psi}{\psi_{0}} ,\,\,\, \tilde{V}\left( x\right) =\frac{V\left(x \right)}{E_{0}},
\end{equation}  
where $x_{0}=\frac{\pi\hbar^{2}\sqrt{\delta g}}{\sqrt{2}mg^{\frac{3}{2}}}$,  $t_{0}=\frac{\pi^{2}\hbar^{2}\delta g}{2mg^{2}}$, $E_{0}=\frac{\hbar^{2}}{mx_{0}^{2}}=\frac{\hbar}{t_{0}}$, and $\psi_{0}=\frac{\sqrt{2m}g^{\frac{3}{2}}}{\pi\hbar\delta g}$.
In terms of these re-scaled parameters, Eq.(1) can be written as
\begin{equation}
i\frac{\partial\psi}{\partial t}=-\frac{1}{2}\frac{\partial^{2}\psi}{\partial x^{2}}+V\left(x \right)\psi+|\psi|^{2}\psi-|\psi|\psi. 
\label{Eq2}
\end{equation} 
Here the tildes are omitted. For a stationary solution $\phi\left(x \right) $, we substitute $\psi \left(x,t \right)=\phi \left(x \right) e^{-i\mu t} $ in Eq.(\ref{Eq2}) and obtain the following equation.
\begin{equation}
\mu\phi=-\frac{1}{2}\frac{d^{2}\phi}{dx^{2}}+V\left(x \right) +\phi^{3}-\phi^{2}.
\label{Eq3}
\end{equation} 
Here $\mu$ is the chemical potential.
In quasi one-dimensional (Q1D) geometry, the effect of trapping on the BECs can be ignored near the center of the trap\cite{q2}. Under this circumstance, Eq.(\ref{Eq3}) gives rise to a family of exact QD solutions \cite{r10} 
\begin{equation}
\phi_{d}\left(x \right) =-\frac{3\mu}{1+\sqrt{1+\frac{9\mu}{2}}\cosh\left(\sqrt{-2\mu}x \right) }
\label{Eq4}
\end{equation} 
for $0<-\mu<\frac{2}{9}$.  The solution in  Eq.(\ref{Eq4}) can also be treated as a QD in free space for $ \Omega >0$ and $\delta g>0$. In this limit, the beyond mean-field effect plays a significant role.  However, one can realize the effect of mean-field interaction  by appropriately tuning the intra- and inter- atomic interactions such that $\delta g<0$ \cite{y1,y2}.  This is, in principle, possible if both the intra- and inter- atomic interaction are attractive and  $|g_{12}|>g$ . In this case, $ \Omega$ can be negligible and the solution of the Eq.(\ref{Eq3}) can approximately be written as  
\begin{equation}
\phi_{l}\left(x \right) \simeq \left( \sqrt{-2\mu} \right) {\rm sech}\left( \sqrt{-2\mu}x\right).
\label{Eq5} 
\end{equation}
This describes a spatially localized solution of attractive BECs in quasi-one-dimensional geometry.

In Fig.1, we plot density profile $\rho(=|\psi|^{2} )$ for different values  of $\mu$. We see that as $\mu\rightarrow-\frac{2}{9}$  the top of a density profile becomes gradually flat  since the effect of quantum fluctuation dominates over the mean-field interaction as $\delta g>0$ (Top panel). In this case,  the effect of quantum fluctuation is maximum. The density profile starts to change  from flat-top to the sharp-top  as $\mu\rightarrow -0.125$ (bottom panel, red curve). Interestingly, the shape of density profiles for $\delta g>0$ and $\delta g<0$ nearly resembles if  $\mu>-0.125$ (bottom panel). This may imply that the  effects of mean-field nonlinear attractive  interaction starts to dominate over the quantum fluctuation as $\mu >-0.125$. It is worth mentioning that a proper modification of  transverse confinement and an appropriate tuning of atomic interaction are required for the change of $\delta g$ from positive ($\delta g>0$)  to negative 
($\delta g<0$) value \cite{n3,sr1,sr2}. In this context, we note that the width of a free-space QD given in Eq.(\ref{Eq4}) can be squeezed in presence of harmonic trap \cite{h1}.  
\begin{figure}[t] 
\includegraphics[scale=0.6]{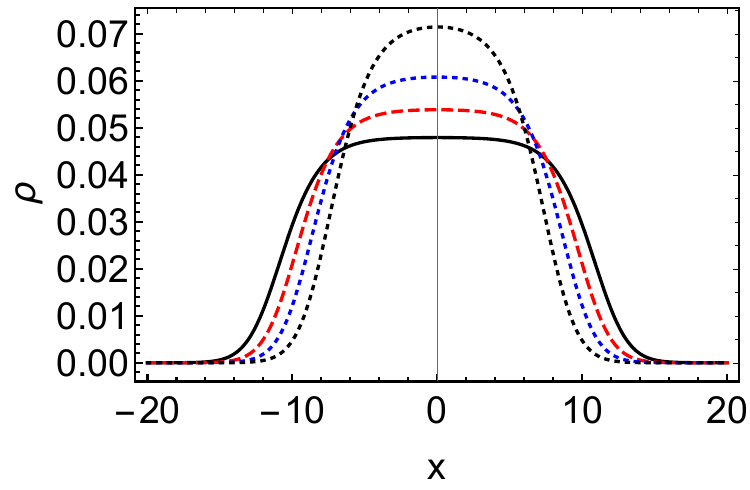} 
\includegraphics[scale=0.6]{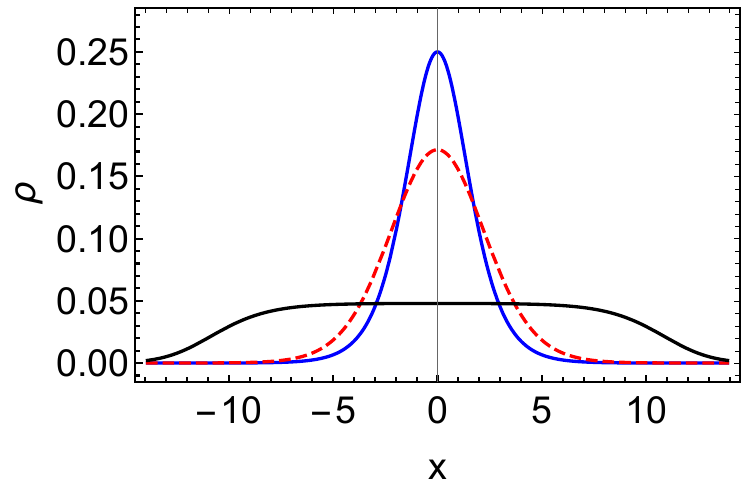} 
\caption{ Top panel: Probability density $\rho=|\phi_{d}|^{2}$ in the position space with $\delta g>0$ for different values of chemical potential $\mu =-0.2222221111111$ (black solid), $-0.2222217$ (red dashed), $-0.22222$ (blue dotted), and $-0.22221$ (black dotted) respectively. Bottom panel: Black and red-dashed curves give density profiles($\rho=|\phi_{d}|^{2}$) with $\delta g>0$ for $\mu =-0.2222221111111$, and $-0.125$ respectively. The  blue curve represents density profile $\rho=|\phi_{l}|^{2}$ given in Eq.(\ref{Eq5}) for $\mu=-0.125 $ and $\delta g<0$.  Here, both $\phi_{d}$ and $\phi_{l}$ are normalized to $1$. }
\end{figure} 
\section{Measurement of information of quantum droplets}   
The theoretical formulation to measure information introduced by C. Shannon in 1948, called Shannon entropy, is considered as the foundation of information theory. Subsequently, various methods, namely, Fisher information, R\'enyi entropy, and Tsallis entropy, for information measures have been developed over the years. These information theoretic approaches are used widely to understand properties of several physical systems \cite{r06,r07}. In the following, we use these four information measures in BECs during the formation of QDs.  
\subsection{ Shannon entropy} 
We have seen that the probability density distribution $\left( \rho\left(x \right)\right)$ of a QD changes  with the chemical potential. To detect the change in global character of the distribution, we first consider Shannon entropy.  In position $\left(S_{\rho} \right) $ and momentum  $\left(S_{\gamma} \right) $ spaces the Shannon entropies are defined by \cite{r12}  
\begin{equation}
 S_{\rho}=-\int\rho\left(x \right)ln\rho\left(x \right)dx,  
\end{equation} 
\begin{equation}
S_{\gamma}=-\int\gamma\left(p\right)ln\gamma\left(p \right)dp.  
\end{equation}
 Here $\rho\left(x \right) $ and $\gamma\left(p \right) $  stand for density distributions in position and momentum spaces corresponding to the wave function $\phi_{d}$ and its Fourier transform($\phi'_{d}$). Understandably, both $\phi_{d}$ and $\phi'_{d}$ are normalized to $1$. 
 The  entropies  $\left(S _{\rho}\right) $ and $\left(S _{\gamma}\right) $  satisfy a relation, often called Bialynicki-Birula and Mycielski $\left(BBM \right) $ inequality \cite{r13,r14}. It is defined as 
\begin{equation}
 S_{t}=S_{\rho}+S_{\gamma}\geq D \left(1+\ln\pi \right). 
\end{equation} 
Here, $D$ is the dimension of the system. This inequality is considered as a stronger version of uncertainty relation. It shows reciprocity between the position- and momentum- space entropies such that a higher value of $S_{\rho}$ is associated with a lower value of $S_{\gamma}$. 
 
In Fig.2, we plot the variation of Shannon entropy with chemical potential ($\mu $). It is seen that, for $-0.125<\mu< -0.200$, the Shannon entropy in position space increases as $\mu$ increases(top panel).  In this range of $\mu$, we get sharp top QDs. As $\mu\rightarrow -2/9$ the entropy grows abruptly due the appearance of flat top QDs. Interestingly, around $\mu \approx -0.20$ a transition occurs from sharp to flat top QDs  resulting a minimum in the entropy.  This implies that  the  onset of QD phase can be identified  from gradual  decrease of $S_\rho$ while the appearance of flat top QDs can be inferred from the abrupt increase of $S_\rho$.  As expected,  the Shannon entropy in momentum space shows a behavior which is exactly opposite to that found in coordinate space(bottom panel).   { We see that the Shannon entropy $(S_\rho)$ in momentum space becomes  negative for the flat-top QDs.  This is due to the fact that the momentum space distributions corresponding to flat-top QDs in coordinate space are highly localized.  In this case, the peak of the QD probability density constructed from a normalized wave function becomes greater than $1$ and thus  results $S_\gamma <1$. It may be noted that  the  occurrence of negative Shannon entropy  either  in coordinate or in momentum space has also been reported in a plenty  of studies  as a result of larger probability density ($\rho >>1$) at some points where $-\rho \ln \rho<0$ \cite{rn14a,rn14b,rn14c,rn14d,rn14e,rn14f}.} 
\begin{figure}[t] 
\includegraphics[scale=0.6]{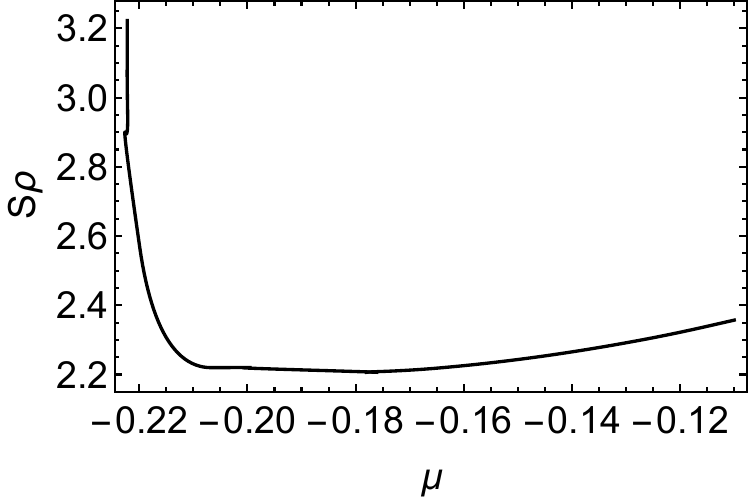}
\includegraphics[scale=0.65]{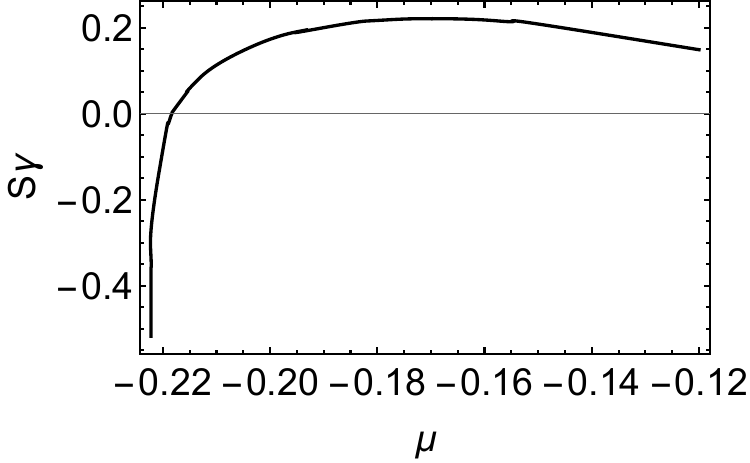}
\caption{Variations of Shannon entropy  in position space (top panel) and momentum space (bottom panel) with $\mu$.}
\end{figure} 
\begin{table}
\begin{tabular}{|l|l|l|l|}
\hline
$\mu$ & $S_{\rho}$& $S_{\gamma}$&$S_{\rho}+S_{\gamma}$\\
\hline
 -0.12&2.3220&0.1487&3.7710\\
 \hline
 -0.15&2.2432&0.2084&2.4516\\
 \hline
 -0.200&2.2201&0.1730&2.3931\\
 \hline
 -0.215&2.3066&0.1723&2.4789\\
 \hline
 -0.22219&2.8262&-0.3441&2.4821\\
 \hline
 -0.22221&2.9046&-0.3579&2.5467\\
 \hline
 -0.22222&3.0308&-0.3802&2.6506\\
 \hline
 -0.2222217&3.1278&-0.4904&2.6374\\
 \hline
 -0.2222221-$\epsilon$&3.2226&-0.5170&2.7056\\
 \hline
 \end{tabular}
\caption{Shannon entropy and entropic uncertainty for quantum droplets. Here $\epsilon=111.111\times 10^{-10}$.}
\end{table}

It is an interesting curiosity to check whether $S_{\rho}$ and $S_{\gamma}$ satisfy the stronger version of uncertainty relation given in Eq.(9) for $D=1$. In order of this, we calculate $S_{\rho}+S_{\gamma}$  for different values of $\mu$ and display the results in Table I. Clearly, the sum of $S_{\rho}$ and $S_{\gamma}$ is always greater than $2.145$. 
	
\subsection{R\'enyi entropy }
A. R\'enyi has proposed an extended version of  Shannon entropy  often, termed as R\'{e}nyi entropy. It is a generalized form of Shannon entropy  containing  a parameter $\left( \alpha\right) $. The sensitivities of  measurements corresponding to different probability distributions depend on the value of $\alpha$. For a probability distribution $\rho\left(x \right) $, the  R\'enyi entropy in coordinate space \cite{r21,r22,r23} is  defined by 
\begin{equation}
R^{\left(\alpha \right) }_{\rho} =\frac{1}{1-\alpha}ln\left[ \int\rho^{\alpha}\left(x \right)dx \right], \,\,\,\alpha\neq1.
\end{equation}	 
If $\gamma\left(p \right)$ stands for the probability distribution in momentum space corresponding to  $\rho\left(x \right) $ then the R\'enyi entropy in momentum space is given by  
\begin{equation}
R^{\left(\alpha \right) }_{\gamma} =\frac{1}{1-\alpha}ln\left[ \int\gamma^{\alpha}\left(p \right)dp \right]. 
\end{equation}	
These two entropies satisfied the following  uncertainty relation\cite{r24}  
\begin{equation} 
R^{\left(\alpha \right) }_{\rho}+R^{\left( \alpha\right) }_{\gamma}\geq -\frac{1}{2}\frac{1}{1-\alpha}ln\left(\frac{\alpha}{\pi} \right). 
\end{equation} 
{It is interesting to note that the R\'enyi index $\alpha$  can take any value from zero to infinity.  For different values of $\alpha$,  the Renyi entropy can have different interpretations. Recently, it is shown that  Renyi entropy for  (i) $\alpha\rightarrow 0$, (ii) $\alpha\rightarrow 1$ and (ii) $\alpha=2$ correspond to   Hartley entropy(max-entropy),  Shannon entropy and collision entropy  respectively.  The R\'enyi entropy with a larger value of $\alpha$ highlights events with higher probabilities while it takes into account events with finite probabilities in rather equal manner for a smaller value of $\alpha$ \cite{r23a} }.
		
In order to verify the results predicted by Shannon entropy  for sharp-top  and flat-top QDs, we calculate R\'e nyi entropy  with $\mu$ for different values $\alpha$ both in coordinate and momentum  spaces. The result is displayed in Fig.3.  {It clearly shows that  the R\'enyi entropy deviates from Shannon entropy as $\alpha$ increases. However, the trend of entropy variation for all $\alpha$ values  remains same. For example, the value of $R^{(\alpha)}_{\rho}$ starts to diverse for $\mu > -0.22$ and  becomes minimum for $-0.125<\mu<-0.200$. It again starts to increase as $\mu < -0.125$.} Like the Shannon entropy, the R\'enyi entropy in momentum space exhibits a trend opposite to that in coordinate space for any value of $\alpha$. {Unlike Shannon entropy, the R\'enyi entropy coinciding with Hartley entropy for $\alpha=2$  can be negative both for flat-top and sharp-top QDs. However, it remains positive during the transition from flat-top to sharp-top QDs.  As $\alpha$ increases, the R\'enyi entropy tends to highlight the event with larger probability and, this ultimately results negative R\'enyi entropy even during the transition from sharp-top to flat-top QDs.}
		
With a view  to check the uncertainties in the measurements of R\'enyi entropy in position and momentum spaces, we take two different values of $\alpha$, namely, $\alpha=2$ and $4$.  The variation of the so-called R\'enyi uncertainty is shown in Table II. We see that the uncertainty is minimum during the transition from sharp to flat-top QDs and it increases  for both sharp-top  and flat-top QDs. However, the value of R\'enyi uncertainty for a sharp top QD is always smaller than that of a flat-top QD.  Like the uncertainty relation for Shannon entropy, the  R\'enyi uncertainty relation is found to be satisfied for $0<-\mu<\frac{2}{9}$.
		
\begin{table}
\begin{tabular}{|l|l|l|l|l|l|l|}
\hline 
$\mu$ & $R^{(2)}_{\rho}$& $R^{(2)}_{\gamma}$&$R^{(2)}_{\rho\gamma} $& $R^{(4)}_{\rho}$& $R^{(4)}_{\gamma}$&$R^{(4)}_{\rho \gamma} $\\
\hline
-0.12&2.1461&-0.0216&2.1245&2.0198&-0.1469&1.8729\\
\hline
-0.15&2.0710&-0.0435&2.0275&1.9470&-0.0793&1.8677\\
\hline
-0.200&2.0619&-0.0199&2.0423&1.9467&-0.0947&1.8520\\
\hline
-0.215&2.1626&-0.0994&2.0632&2.0571&-0.2102&1.8469\\
\hline
-0.2215&2.4226&-0.34432&2.0783&2.3372&-0.4633&1.8739\\
\hline
-0.22219&2.7388&-0.6018&2.1370&2.6752&-0.7418&1.9334\\
\hline
-0.22221&2.8239&-0.6408&2.1831&2.7656&-0.7942&1.9714\\
\hline
-0.22222&2.9601&-0.6959&2.2642&2.9094&-0.8744&2.0350\\
\hline 
-0.2222217&2.0639&-0.8225&2.2414&3.0184&-0.9968&2.0216\\
\hline 
-0.2222221-$\epsilon$&3.1648&-0.8686&2.2962&3.1238&-1.0520&2.0718\\
\hline
\end{tabular}
\caption{R\'enyi entropy and entropic uncertainty of quantum droplet for $\alpha=2$ and $4$. Here  $R^{(\alpha)}_{\rho \gamma}=R^{(\alpha)}_{\rho}+ R^{(\alpha)}_{\gamma} $ and $\epsilon=111.111\times 10^{-10}$.}
\end{table} 
\begin{figure} 
\includegraphics[scale=0.6]{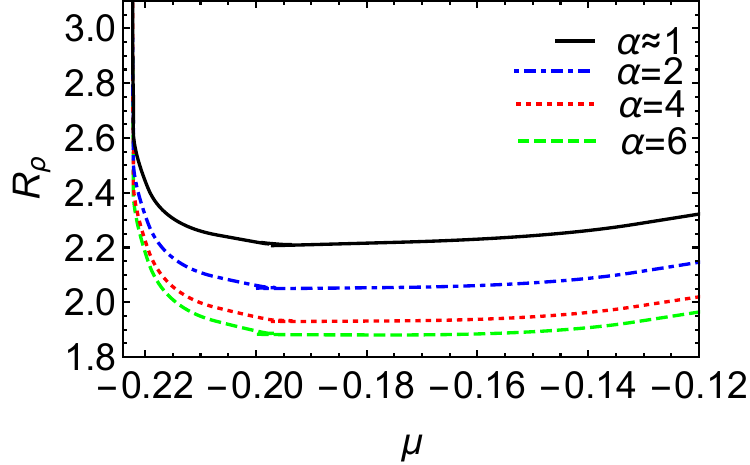}
\includegraphics[scale=0.6]{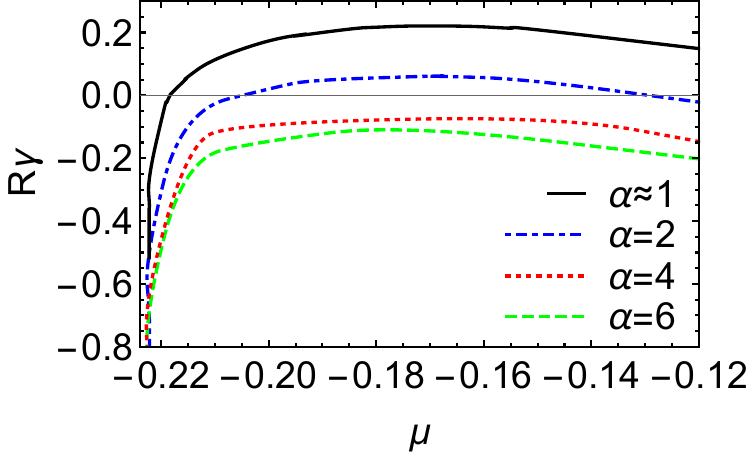}
\caption{Variations of R\'enyi entropy in position space(top panel) and momentum space(bottom panel) with chemical potential for different values of $\alpha$.}
\end{figure}  
\subsection{Tsallis Entropy}
 There is another important information measure, called Tsallis entropy,  which also gives Shannon entropy as a limiting case.  The Tsallis entropy for the probability distributions, $\rho\left(x \right) $ in coordinate space and $\gamma\left(p \right) $ in momentum space are defined as \cite{r05,r04,r03,r25} 
\begin{equation} 
T^{\left( q\right) }_{ \rho }=\frac{1}{q-1} \left[1-\int\rho^{q}\left(x \right)dx  \right] 
\end{equation} 
		and  
\begin{equation} 
	T^{\left(q \right) }_{\gamma  }=\frac{1}{q-1} \left[1-\int\gamma^{q}\left(p \right)dp  \right] 
\end{equation} 
respectively. Here $q$ is a real parameter which we call it as Tsallis entropy index. The Tsallis entropy tends to Shannon entropy as $q\rightarrow 1 $. Understandably,  sensitivity of the Tsallis entropic measure corresponding  to different probability distributions depends on  $q$.

\begin{figure}[t] 
\includegraphics[scale=0.36]{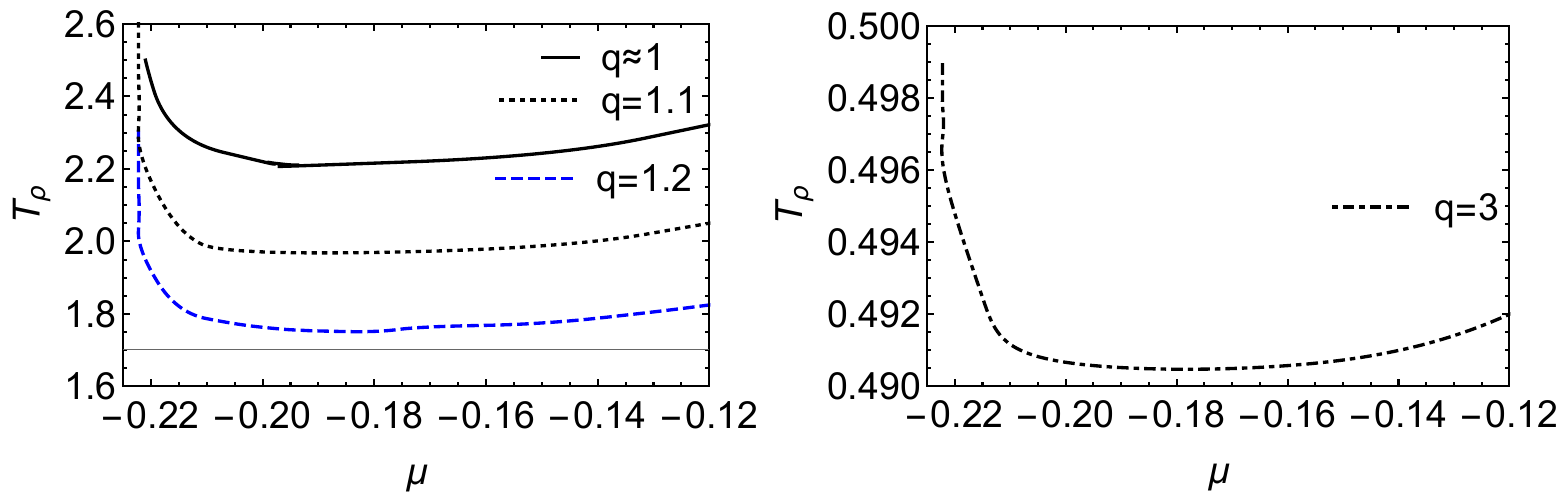}
\includegraphics[scale=0.35]{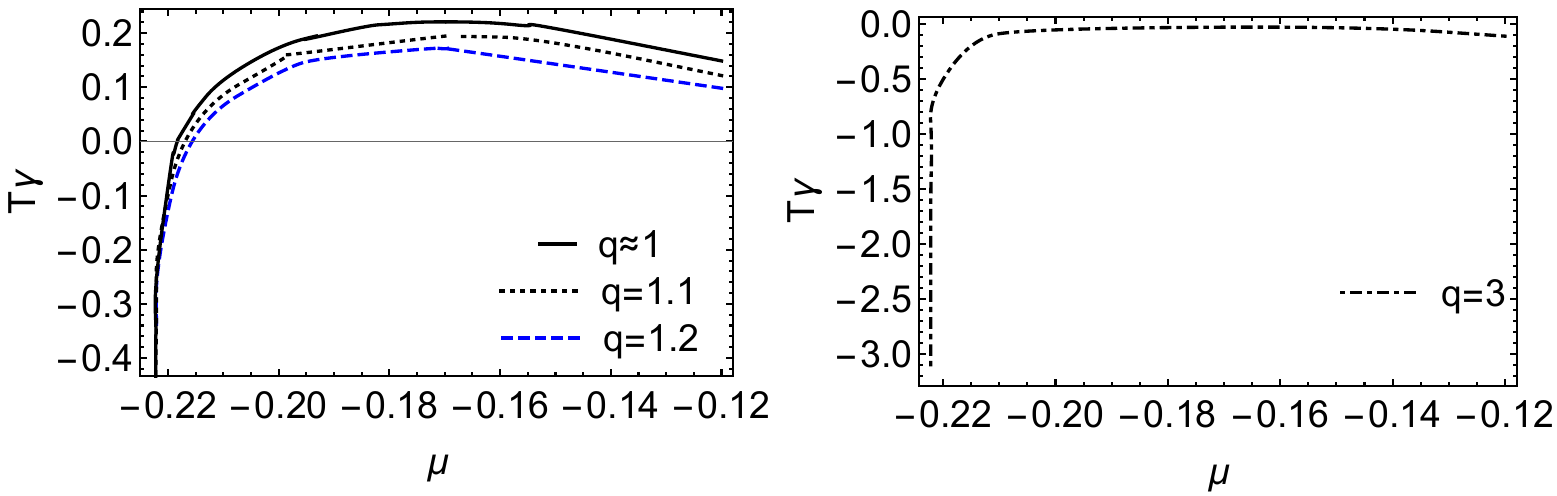}
\caption{Variations of Tsallis entropy with chemical potential of a QD in coordinate-
(top panel) and momentum-space(bottom panel) for different values of $q$. } 
\label{fig4}
\end{figure}

In Fig.\ref{fig4}, we plot Tsallis entropy in coordinate ($T^{(q)}_{\rho}$) and momentum ($T^{(q)}_{\gamma}$) spaces for different values of $q$. {We see that, like Shannon and R\'enyi entropies,   $T^{(q)}_{\rho}$  for $\mu\rightarrow -\frac{2}{9}$ (flat-top QD) and $\mu > -0.125$ (sharp-top QD) in coordinate space increases. During the transition between these two cases, the Tsallis entropy becomes minimum. The trend of variation remains same  for $q=1$ and $q>1$. We note  that this entropy is very sensitive with the index $q$. As $q$  becomes slightly greater than $1$, the Tsallis entropy deviates appreciably from the Shannon entropy ($q\approx 1$). For a relatively larger value of $q$, Tsallis entropy in momentum space becomes negative for both Sharp-top and flat-top QDs. Thus the value of Tsallis entropy for $q>1$ differs from Shannon entropy ($S_\gamma$) in the sense that $S_\gamma$ is negative only for flat-top QDs.  It is worth mentioning  that the sum ($T^{(q)}_{\rho\gamma}$) of Tsallis entropies  in coordinate ($T^{(q)}_{\rho}$) and momentum ($T^{(q)}_{\gamma}$) spaces is positive for sharp-top QDs and it becomes  negative as the sharpness of a QD top deceases.}
 
\begin{table}
\begin{tabular}{|l|l|l|l|l|l|l|}
\hline 
$-\mu$ & $T^{(3)}_{\rho}$& $T^{(3)}_{\gamma}$&$T^{(3)}_{\rho\gamma}$& $T^{(5)}_{\rho}$& $T^{(5)}_{\gamma}$&$T^{(5)}_{\rho\gamma}$\\
\hline
0.12&0.4920&-0.1106&0.3814&0.2499&-0.1612&0.0887\\
\hline
0.15&0.4907&-0.0343&0.4564&0.2499&-0.1394&0.1105\\
\hline
0.195&0.4904&-0.0372&0.4532&0.2499&-0.1378&0.1121\\
\hline
0.200&0.4907&-0.0539&0.4367&0.2499&-0.1613&0.0886\\
\hline
0.2215&0.4956&-0.6584&-0.16279&0.2500&-1.5416&-1.2916\\
\hline 
0.2221&0.4970&-1.0835&-0.5864&0.2500&-3.1772&-2.9272\\
\hline
0.22219&0.4977&-1.4998&-1.0021&0.2500&-5.2625&-5.0125\\
\hline
0.22221&0.4981&-1.7027&-1.2046&0.2500&-6.6102&-6.3602\\
\hline
0.22222&0.4986&-2.0458&-1.573&0.2500&-9.3540&-9.1040\\
\hline 
0.2222217&0.4988&-2.7645&-2.2656&0.2500&-15.3647&-15.1147\\
\hline 
0.2222221+$\epsilon$&0.4991&-3.1210&-2.6219&0.2500&-19.4155&-19.1655\\
\hline
\end{tabular} 
\caption{Tsallis entropy and $T^{(q)}_{\rho\gamma}$  for $q=3$ and $5$. Here $T^{(q)}_{\rho\gamma}=T^{(q)}_{\rho}+ T^{(q)}_{\gamma}$ and $\epsilon=111.111\times 10^{-10}$.}
\end{table} 
\subsection{Fisher information}  
As pointed out, the Fisher information is sensitive to local rearrangement of the density distribution and thus it gives a local information measure. With a view  to extract  information due to the local change of probability distribution during the formation of QDs, we consider Fisher information both in coordinate ($F_{\rho}$) and momentum($F_{\gamma}$) spaces which are defined by \cite{r58} 
\begin{equation}
 F_{\rho}=\int\frac{1}{\rho\left(x \right) }\left[ \frac{d\rho\left(x \right) }{dx}\right] ^{2}dx,
\end{equation}  
  \begin{equation}
 	F_{\gamma}=\int\frac{1}{\gamma\left(p \right) }\left[ \frac{d\gamma\left(p \right) }{dp}\right] ^{2}dp
 \end{equation} 
 respectively. The Fisher informations $F_{\rho}$ and $F_{\gamma}$ satisfy the so-called {Cramer-Rao (CR)} inequality. For an one-dimensional(1D) system, the CR inequality \cite{r59,r60} is given by 
\begin{equation}
 F_{\rho}F_{\gamma}\geq 4.
\label{Eq11}
 \end{equation}
This gives a lower bound to the product of Fisher information in position and momentum spaces. {We note that the CR inequality holds good if one of the  wave functions between position and momentum spaces is real \cite{r61}.  For arbitrary complex wave functions, the CR inequality is not valid.   There are  a plenty of examples, namely, Lane-Emden oscillator, Caldirola-Kanai oscillator \cite{n4,n6}, time evolution of free particle \cite{r61} where $F_{\rho}F_{\gamma}< 4$.  Also the Hall's conjecture \cite{q1} demands that the inequality in Eq.(\ref{Eq11})  is true for pure states. Therefore, the validity of the CR inequality is not a necessary condition but a sufficient condition \cite{r61,q1}.}  
\begin{table}
\begin{tabular}{|l|l|l|l|l|}
\hline
$\mu$ & $F_{\rho}$& $F_{\gamma}$&$F_{\rho} F_{\gamma}$&$F_\rho N_\rho$\\
\hline
-0.12&0.1665&12.7814&2.1281&17.3118\\
\hline
-0.15&0.1943&11.3141&2.1983&17.2522\\
\hline
-0.200&0.2020&12.4405&2.5130&17.1260\\
\hline
-0.2215&0.1124&31.4202&3.5316&18.0153\\
\hline
-0.22215&0.0796&45.2155&3.5991&19.7357\\
\hline
-0.22219&0.0719&81.4983&5.8598&20.4892\\
\hline
-0.22221&0.0644&72.8636&4.6924&21.4667\\
\hline
-0.222221&0.0515&82.3974&4.2434&24.0087\\
\hline 
-0.2222217&0.0480&141.5720&6.7954&25.0009\\
\hline
-0.2222221-$\epsilon$&0.0426&150.3520&6.4050&26.8541\\
\hline
\end{tabular}
\caption{Fisher information and Cramer-Rao inequality for different values of  $ \mu$. {The Shannon-Fisher complexity ($C_\rho$) is given in the last column}. Here $\epsilon=111.111\times10^{-10}$.}
\end{table}  

In Fig.5, we plot $F_{\rho}$ and $F_{\gamma}$ for different chemical potentials both in coordinate (top panel) and momentum  (bottom panel) spaces. We see that, unlike Shannon entropy, the $F_{\rho}$ decreases for both sharp-top and flat top QDs  in coordinate space. However, the change of $F_{\rho}$ with $\mu$ is relatively faster than that of Shannon entropy. We observe  abrupt decrease of $F_{\rho}$ for flat top QDs ($\mu\rightarrow -\frac{2}{9}$) and gradual decrease of $F_{\rho}$ for  sharp-top QDs ($\mu>-0.200$). In between of this two limiting cases, $F_{\rho}$ attains its maximum value. The Fisher information in momentum space shows a trend opposite to that in coordinate space. The values of uncertainty relations for Fisher information in coordinate and momentum spaces are shown in Table IV. We see that the product of  Fisher information of flat-top QDs satisfy CR inequality while {it fails to satisfy the same  for sharp-top QDs ($\mu\geq-0.22219$). Another important inequality  due to Stam\cite{r20} involving Shannon entropy ($S_\rho$) and Fisher information ($F_\rho$), called Shannon-Fisher complexity($C_\rho$), is given by\cite{r71} 
\begin{eqnarray}
F_\rho N_\rho\geq 2\pi\,e
\label{eq18}
\end{eqnarray}
with $N_\rho=\exp(2S_\rho$). We see that this inequality is satisfied  for both sharp-top and flat-top QDs, and the value of $C_\rho$ increases as the QD changes its shape from sharp-top to flat-top (Table IV).}

\begin{figure}[t] 
\includegraphics[scale=0.52]{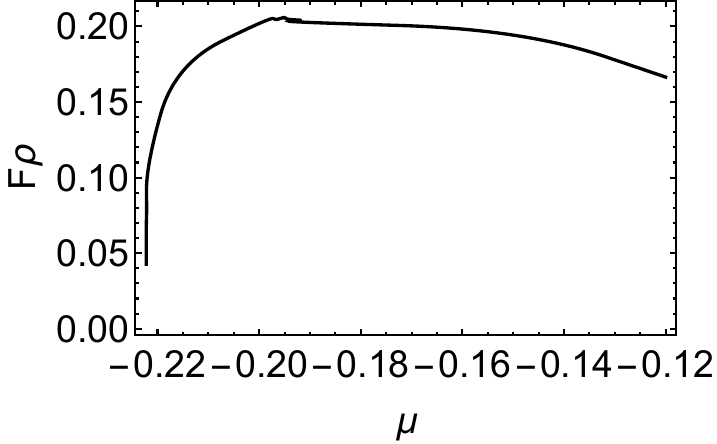}
\includegraphics[scale=0.5]{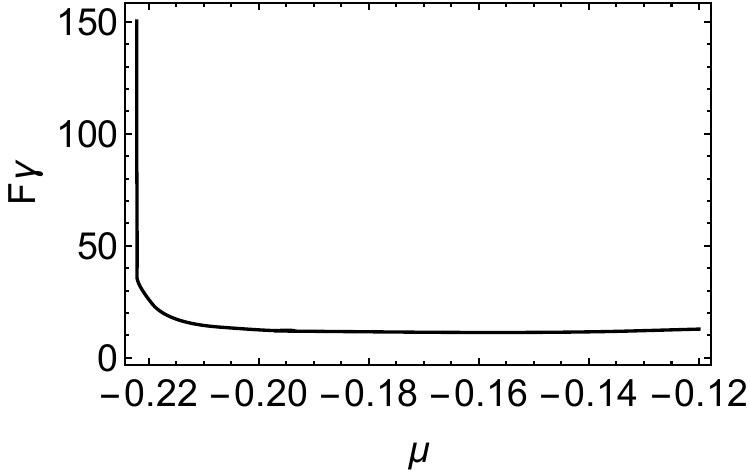}
\caption{Variations of Fisher information in position space (top panel) and momentum space (bottom panel) with chemical potential of a quantum droplet.}
\end{figure} 
\begin{figure}[t] 
\includegraphics[scale=0.35]{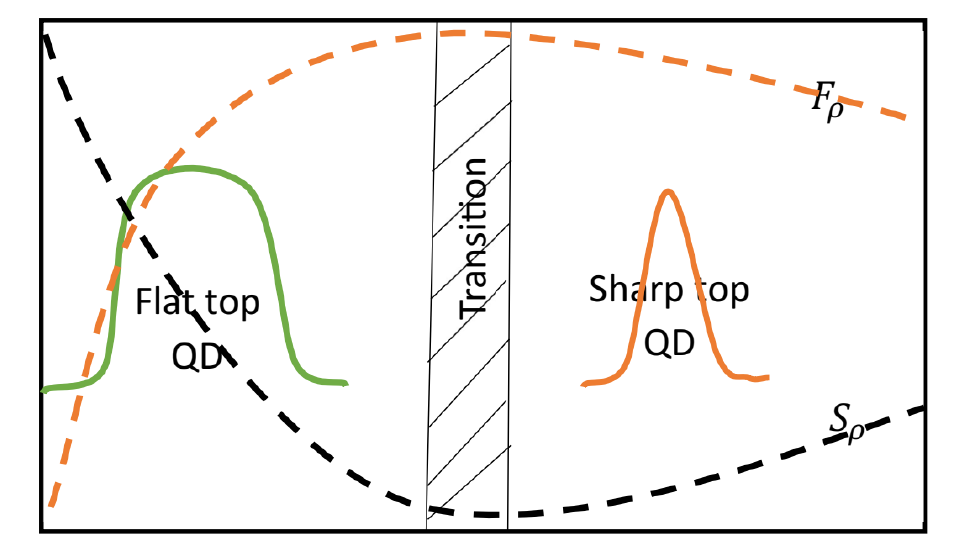}
\caption{Schematic diagram of different for  flat-top to sharp-top  quantum droplets with entropy change.} 
\label{fig6}
\end{figure}

{In Fig.\ref{fig6}, we  show a schematic diagram to visualize the obtained results. It shows that the changes of $F_\rho$ and $S_\rho$ for flat-top QDs are faster than those for sharp-top QDs. A region of transition between these two types of QDs is identified from the occurrence  of minimum and maximum in $S_\rho$ and $F_\rho$ respectively.}
\section{ Conclusion} 
 Information theoretic approach provides a powerful framework for understanding and quantifying information in various systems. This approach is based on the nature of  probability  distribution  and thus any change in the probability distribution is reflected in the information entropic measures. However, sensitivity of  measurement determines the accuracy of information. There are several entropic measures formulae to extract information from various systems. In this work, we have used four important and popular information measures, namely, Shannon entropy, Fisher information, R\'enyi and Tsallis entropies to study quantum droplet phase in Bose Einstein condensates. 
 
 We have seen that the information theoretic measure provides  a straight forward approach to characterize quantum droplets associated with the change of chemical potential in Bose Einstein condensation. At onset of quantum droplet phase, chemical potential is relatively large and the top of density distribution of a quantum droplet(QD) is sharp. The top of a QD becomes gradually flat due to quantum fluctuation  as the the chemical potential decrease. We know that the information theoretical measure can detect global or local shape change of a distribution and thus can identify the onset of quantum droplet formation.  Particularly, in coordinate space, the value of Shannon entropy decreases during the onset of QD phase and then attains a minimum value at the point where a transition occurs from sharp to flat tops QDs. It increases abruptly with the decrease of chemical potential for the flat top QDs. We have shown that the Fisher information exhibits a behaviour exactly opposite to that of Shannon entropy.  For example,  Fisher information increases for both flat and sharp top quantum droplets. Also during the transition from sharp to flat top QDs Fisher information attains a maximum value.  It is seen that the results obtained from other two information measures, namely, R\'enyi and Tsallis entropies are consistent with those predicted by Shannon entropy and Fisher information. We conclude by noting that the stable quantum droplets permitted by perturbative Lee-Haung-Yang(LHY) model in weakly interactive ultracold atomic system can  efficiently be detected with the help of information theoretic measurements.

\subsection*{Acknowledgment:}  
S. Siddik would like to thank "West Bengal Higher Education Department" for providing Swami Vivekananda Merit Cum Means Scholarship.

\end{document}